\makeatletter
\newcommand{\dontusepackage}[2][]{%
  \@namedef{ver@#2.sty}{9999/12/31}%
  \@namedef{opt@#2.sty}{#1}}
\makeatother
\dontusepackage{subfigure}

\documentclass[]{article}

\usepackage{lmodern}
\usepackage{amssymb,amsmath}
\usepackage{ifxetex,ifluatex}
\usepackage[usenames,dvipsnames]{color}
\usepackage{fixltx2e} 
\ifnum 0\ifxetex 1\fi\ifluatex 1\fi=0 
  \usepackage[T1]{fontenc}
  \usepackage[utf8]{inputenc}
\else 
  \ifxetex
    \usepackage{mathspec}
    \usepackage{xltxtra,xunicode}
  \else
    \usepackage{fontspec}
  \fi
  \defaultfontfeatures{Mapping=tex-text,Scale=MatchLowercase}
  
\fi
\IfFileExists{upquote.sty}{\usepackage{upquote}}{}
\IfFileExists{microtype.sty}{%
\usepackage{microtype}
\UseMicrotypeSet[protrusion]{basicmath} 
}{}
\usepackage[margin=1.0in,bottom=1.5in]{geometry}
\usepackage[square,numbers,sort&compress]{natbib}
\usepackage{listings}
\usepackage{graphicx}
\makeatletter
\def\maxwidth{\ifdim\Gin@nat@width>\linewidth\linewidth\else\Gin@nat@width\fi}
\def\maxheight{\ifdim\Gin@nat@height>\textheight\textheight\else\Gin@nat@height\fi}
\makeatother
\setkeys{Gin}{width=\maxwidth,height=\maxheight,keepaspectratio}
\usepackage{caption}
\usepackage{float}
\usepackage{subfig}
\ifxetex
  \usepackage[setpagesize=false, 
              unicode=false, 
              xetex]{hyperref}
\else
  \usepackage[unicode=true]{hyperref}
\fi
\hypersetup{breaklinks=true,
            bookmarks=true,
            pdfauthor={},
            pdftitle={Learned imaging with constraints and uncertainty quantification},
            colorlinks=true,
            citecolor=black,
            urlcolor=blue,
            linkcolor=black,
            pdfborder={0 0 0}}
\usepackage[final]{neurips_2019}
\usepackage{url}
\usepackage{booktabs}
\usepackage{amsfonts}
\usepackage{nicefrac}
\usepackage{microtype}

\def\argmin{\mathop{\rm argmin}}

\newcommand{\minim}{\mathop{\mathrm{minimize}}}

\title{Learned imaging with constraints and uncertainty quantification}
\author{Felix J. Herrmann, Ali Siahkoohi, and Gabrio Rizzuti\\School of
Computational Science and Engineering\\Georgia Institute of
Technology\\\texttt{\{felix.herrmann,\phantom{\ }alisk,\phantom{\ }rizzuti.gabrio\}@gatech.edu}}
\date{}

\begin{document}
\maketitle
\begin{abstract}
We outline new approaches to incorporate ideas from deep learning into
wave-based least-squares imaging. The aim, and main contribution of this
work, is the combination of handcrafted constraints with deep
convolutional neural networks, as a way to harness their remarkable ease
of generating natural images. The mathematical basis underlying our
method is the expectation-maximization framework, where data are divided
in batches and coupled to additional ``latent'' unknowns. These unknowns
are pairs of elements from the original unknown space (but now coupled
to a specific data batch) and network inputs. In this setting, the
neural network controls the similarity between these additional
parameters, acting as a ``center'' variable. The resulting problem
amounts to a maximum-likelihood estimation of the network parameters
when the augmented data model is marginalized over the latent variables.
\end{abstract}

\section{The seismic imaging problem}\label{the-seismic-imaging-problem}

In least-squares imaging, we are interested in inverting the following
inconsistent ill-conditioned linear inverse problem:
\begin{equation}
\minim_x \frac{1}{2}\sum_{i=1}^N\|y_i-A_ix\|_2^2.
\label{eq:LSQR}
\end{equation}
 In this expression, the unknown vector $x$ represents the image,
$y_i,\, i=1, \ldots, N$ the observed data from $N$ source experiments
and $A_i$ the discretized linearized forward operator for the
$i\text{th}$ source experiment. Despite being overdetermined, the above
least-squares imaging problem is challenging. The linear systems $A_i$
are large, expensive to evaluate, and inconsistent because of noise
and/or linearization errors.

As in many inverse problems, solutions of problem~\ref{eq:LSQR} benefit
from adding prior information in the form of penalties or preferentially
in the form of constraints, yielding
\begin{equation}
\minim_x \frac{1}{2}\sum_{i=1}^N\|y_i-A_ix\|_2^2 \quad\text{subject to}\quad x\in\mathcal{C}
\label{eq:constrained-LSQR}
\end{equation}
 with $\mathcal{C}$ representing a single or multiple (convex)
constraint set(s). This approach offers the flexibility to include
multiple handcrafted constraints. Several key issues remain, namely;
\emph{(i)} we can not afford to work with all $N$ experiments when
computing gradients for the above data-misfit objective; \emph{(ii)}
constrained optimization problems converge slowly; \emph{(iii)}
handcrafted priors may not capture complexities of natural images;
\emph{(iv)} it is non-trivial to obtain uncertainty quantification
information.

\section{Stochastic linearized
Bregman}\label{stochastic-linearized-bregman}

To meet the computational challenges that come with solving
problem~\ref{eq:constrained-LSQR} for non-differentiable structure
promoting constraints, such as the $\ell_1$-norm, we solve
problem~\ref{eq:constrained-LSQR} with Bregman iterations for a batch
size of one. The $k\text{th}$ iteration reads
\begin{equation}
\begin{aligned}
\begin{array}{lcl}
    \tilde{x} & \leftarrow & \tilde{x}-t_k A_k^\top(A_kx-y_k)\\
    x & \leftarrow & \mathcal{P}_{\mathcal{C}} (\tilde{x})
\end{array}
\end{aligned}
\label{eq:bregman-iterations}
\end{equation}
 with $A_k^\top$ the adjoint of $A_k$~, where
$A_k \in \left \{ A_i \right \}_{i=1}^{N}$, and
\begin{equation}
\mathcal{P}_{\mathcal{C}} (\tilde{x}) = \argmin_{x} \frac{1}{2} \| x - \tilde{x} \|_2^2 \quad \text{subject to} \quad x \in  \mathcal{C}
\label{proj_card}
\end{equation}
 being the projection onto the (convex) set and
$t_k={\|A_kx-y_k\|_2^2}/{\|A_k^\top(A_kx-y_k)\|_2^2}$ the dynamic
steplength. Contrary to the Iterative Shrinkage Thresholding Algorithm
(ISTA), we iterate on the dual variable $\tilde{x}$. Moreover, to handle
more general situations and to ensure we are for every iteration
feasible (= in the constraint set) we replace sparsity-promoting
thresholding with projections that ensure that each model iterate
remains in the constraint set. As reported in \citet{Witte2019},
iterations~\ref{eq:bregman-iterations} are known to converge fast for
pairs $\{y_k,\,A_k\}$ that are randomly drawn, with replacement, from
iteration to iteration. As such, Equation~\ref{eq:bregman-iterations}
can be interpreted as stochastic gradient descent on the dual variable.

\section{Deep prior with constraints}\label{deep-prior-with-constraints}

Handcrafted priors, encoded in the constraint set $\mathcal{C}$, in
combination with stochastic optimization, where we randomly draw a
different source experiment for each iteration of
Equation~\ref{eq:bregman-iterations}, allow us to create high-fidelity
images by only working with random subsets of the data. While
encouraging, this approach relies on handcrafted priors encoded in the
constraint set $\mathcal{C}$. Motivated by recent successes in machine
learning and deep convolutional networks (CNNs) in particular, we follow
\citet{Veen2018}, \citet{dittmer2018regularization} and
\citet{wu2019parametric} and propose to incorporate CNNs as deep priors
on the model. Compared to handcrafted priors, deep priors defined by
CNNs are less biased since they only require the model to be in the
range of the CNN, which includes natural images and excludes images with
unnatural noise. In its most basic form, this involves solving problems
of the following type \citep{Veen2018}:
\begin{equation}
\minim_w \frac{1}{2}\|y-Ag(z,w)\|_2^2.
\label{eq:deep-prior}
\end{equation}
 In this expression, $g(z,w)$ is a deep CNN parameterized by unknown
weights $w$ and $z\sim\mathrm{N}(0,1)$ is a fixed random vector in the
latent space. In this formulation, we replaced the unknown model by a
neural net. This makes this formulation suitable for situations where we
do not have access to data-image training pairs but where we are looking
for natural images that are in the range of the CNN. In recent work by
\citet{Veen2018}, it is shown that solving problem~\ref{eq:deep-prior}
can lead to good estimates for $x$ via the CNN $g(z,\widehat{w})$ where
$\widehat{w}$ is the minimizer of problem~\ref{eq:deep-prior} highly
suitable for situations where we only have access to data. In this
approach, the parameterization of the network by $w$ for a fixed $z$
plays the role of a non-linear redundant transform.

While using neural nets as strong constraints may offer certain
advantages, there are no guarantees that the model iterates remain
physically feasible, which is a prerequisite if we want to solve
non-linear imaging problems that include physical parameters
\citep{esser2016tvr, peters2018pmf}. Unless we pre-train the network,
early iterations while solving problem~\ref{eq:deep-prior} will be
unfeasible. Moreover, as mentioned by \citet{Veen2018}, results from
solving inverse problems with deep priors may benefit from additional
types of regularization. We accomplish this by combining hard
handcrafted constraints with a weak constraint for the deep prior
resulting in a reformulation of the problem~\ref{eq:deep-prior} into
\begin{equation}
\minim_{x\in \mathcal{C},\,w} \frac{1}{2}\|y-Ax\|_2^2 + \frac{\lambda^2}{2}\|x-g(z,w)\|_2^2.
\label{eq:constrained-deep-prior}
\end{equation}
 In this expression, the deep prior appears as a penalty term weighted
by the trade-off parameter $\lambda>0$. In this weak formulation, $x$ is
a slack variable, which by virtue of the hard constraint will be
feasible throughout the iterations.

The above formulation offers flexibility to impose constraints on the
model that can be relaxed during the iterations as the network is
gradually ``trained''. We can do this by either relaxing the constraint
set (eg. by increasing the size of the TV-norm ball) or by increasing
the trade-off parameter $\lambda$.

\section{Learned imaging via expectation
maximization}\label{learned-imaging-via-expectation-maximization}

So far, we used the neural network to regularize inverse problems
deterministically by selecting a single latent variable $z$ and
optimizing over the network weights initialized by white noise. While
this approach may remove bias related to handcrafted priors, it does not
fully exploit documented capabilities of generative neural nets, which
are capable of generating realizations from a learned distribution.
Herein lies both an opportunity and a challenge when inverse problems
are concerned where the objects of interest are generally not known a
priori. Basically, this leaves us with two options. Either we assume to
have access to an oracle, which in reality means that we have a training
set of images obtained from some (expensive) often unknown imaging
procedure, or we make necessary assumptions on the statistics of real
images. In both cases, the learned priors and inferred posteriors will
be biased by our (limited) understanding of the inversion process,
including its regularization, or by our (limited) understanding of
statistical properties of the unknown e.g.~geostatistics
\citep{mosser2018stochastic}. The latter may lead to perhaps
unreasonable simplifications of the geology while the former may suffer
from remnant imprint of the nullspace of the forward operator and/or
poor choices for the handcrafted and deep priors.

\subsection{Training phase}\label{training-phase}

Contrary to approaches that have appeared in the literature, where the
authors assume to have access to a geological oracle
\citep{mosser2018stochastic} to train a GAN as a prior, we opt to learn
the posterior through inversion deriving from the above combination of
hard handcrafted constraints and weak deep priors with the purpose to
train a network to generate realizations from the posterior. Our
approach is motivated by \citet{han2017alternating} who use the
Expectation Maximization (EM) technique to train a generative model on
sample images. We propose to do the same but now for seismic data
collected from one and the same Earth model.

To arrive at this formulation, we consider each of the $N$ source
experiments with data $y_k$ as separate datasets from which images $x_k$
can in principle be inverted. In other words, contrary to
problem~\ref{eq:LSQR}, we make no assumptions that the $y_k$ come from
one and the same $x$ but rather we consider $n\ll N$ different batches
each with their own $x_k$. Using the these $y_k$, we solve an
unsupervised training problem during which

\begin{itemize}
\item
  $n$ minibatches of observed data, latent, and slack variables are
  paired into tuples $\{y_i,x_i,z_i\}_{i=1}^n$ with the latent variables
  $z_i$'s initialized as zero-centered white Gaussian noise,
  $z_i\sim N(0,I)$. The slack variables $x_i$'s are computed by the
  numerically expensive Bregman iterations, which during each iteration
  work on the randomized source experiment of each minibatch.
\item
  latent variables $z_i$'s are sampled from $p(z_i | x_i, w)$ by running
  $l$ iterations of Stochastic Gradient Langevin Dynamics (SGLD,
  \citet{welling2011bayesian})~(Equation~\ref{eq:Langevin}), where $w$
  is the current estimate of network weights, and $x_i$'s are computed
  with Bregman iterations
  (Equation~\ref{eq:constrained-bregman-iteration}). These iterations
  for the latent variables are warm-started while keeping the network
  weights $w$ fixed. This corresponds to an unsupervised inference step
  where training pairs $\{x_i,z_i\}_{i=1}^n$ are created. Uncertainly in
  the $z_i$'s is accounted for by SGLD iterations
  \citep{han2017alternating, mosser2018stochastic}.
\item
  the network weights are updated using $\{x_i,z_i\}_{i=1}^n$ with a
  supervised learning procedure. During this learning step, the network
  weights are updated by sample averaging the gradients w.r.t. $w$ for
  all $z_i$'s. As stated by \citet{han2017alternating}, we actually
  compute a Monte Carlo average from these samples.
\end{itemize}

By following this semi-supervised learning procedure, we expose the
generative model to uncertainties in the latent variables by drawing
samples from the posterior via Langevin dynamics that involve the
following iterations for the pairs $\{x_i,\, z_i\}_{i=1}^n$
\begin{equation}
z_i\leftarrow z_i-\frac{\varepsilon}{2}\nabla_z\left ( \lambda^2 \|x_i - g(z_i,w))\|_2^2 + \left \| z_i \right \|_2^2  \right ) +\mathcal{N}(0, \varepsilon I)
\label{eq:Langevin}
\end{equation}
 with $\varepsilon$ the steplength. Compared to ordinary gradient
descent,~\ref{eq:Langevin} contains an additional noise term that under
certain conditions allows us to sample from the posterior distribution,
$p(z_i | x_i, w)$. The training samples $x_i$ came from the following
Bregman iterations in the outer loop
\begin{equation}
\begin{aligned}
\begin{array}{lcl}
    \tilde{x}_i & \leftarrow & \tilde{x}_i- t_k  \left ( A_k^\top(A_kx_i-y_k) + \lambda^2 (x_i-g(z_i,w)) \right)\\
    x_i         & \leftarrow & \mathcal{P}_{\mathcal{C}} (\tilde{x}_i).
\end{array}
\end{aligned}
\label{eq:constrained-bregman-iteration}
\end{equation}
 After sampling the latent variables, we update the network weights via
for the $z_i$'s fixed
\begin{equation}
w \leftarrow w -\eta \nabla_w \sum_{i=1}^n\|x_i-g(z_i,w)\|_2^2
\label{eq:constrained-bregman-iterations-b}
\end{equation}
 with $\eta$ steplength for network weights.

Conceptually, the above training procedure corresponds to carrying out
$n$ different inversions for each data set $y_i$ separately. We train
the weights of the network as we converge to the different solutions of
the Bregman iterations for each dataset. As during Elastic-Averaging
Stochastic Gradient Descent
\citep[\citet{chaudhari2016entropy}]{zhang2015deep}, $x_i$'s have room
to deviate from each other when $\lambda$ is not too large. Our approach
differs in the sense that we replaced the center variable by a
generative network.

\section{Example}\label{example}

\begin{figure*}
\centering
\subfloat[\label{figure-1a}]{\includegraphics[width=0.500\hsize]{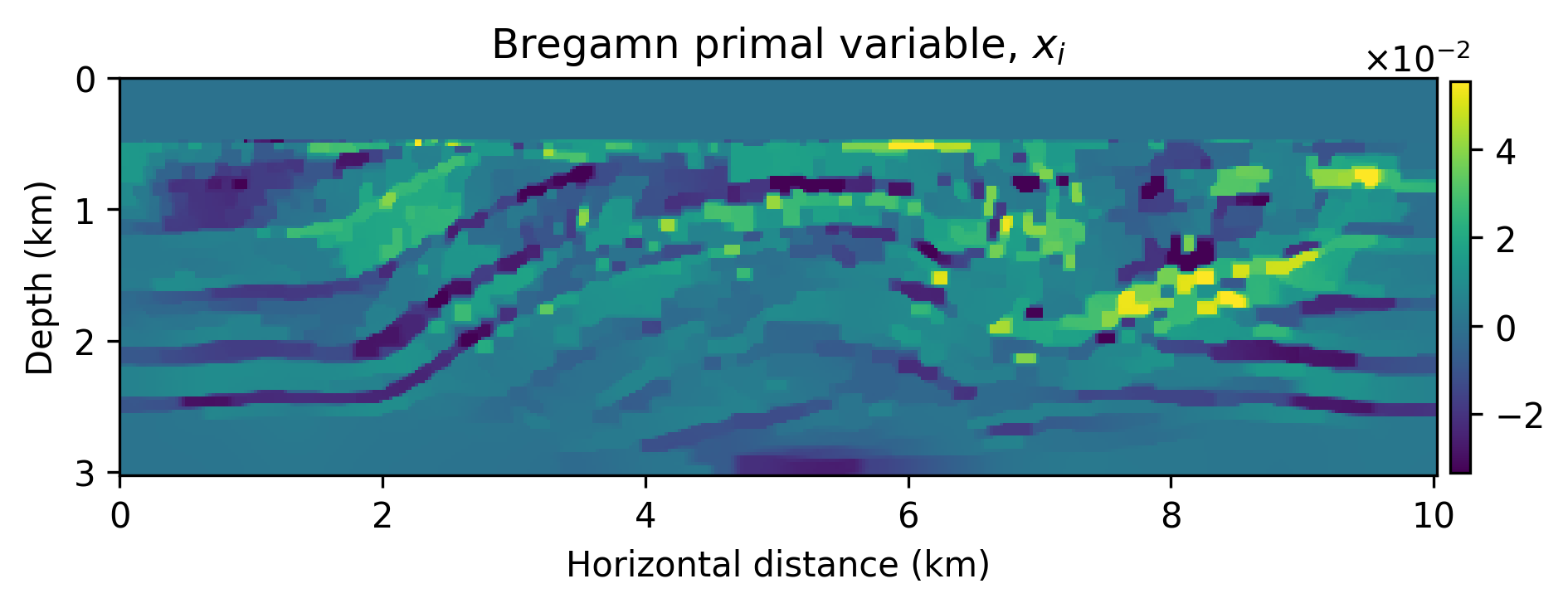}}
\subfloat[\label{figure-1b}]{\includegraphics[width=0.500\hsize]{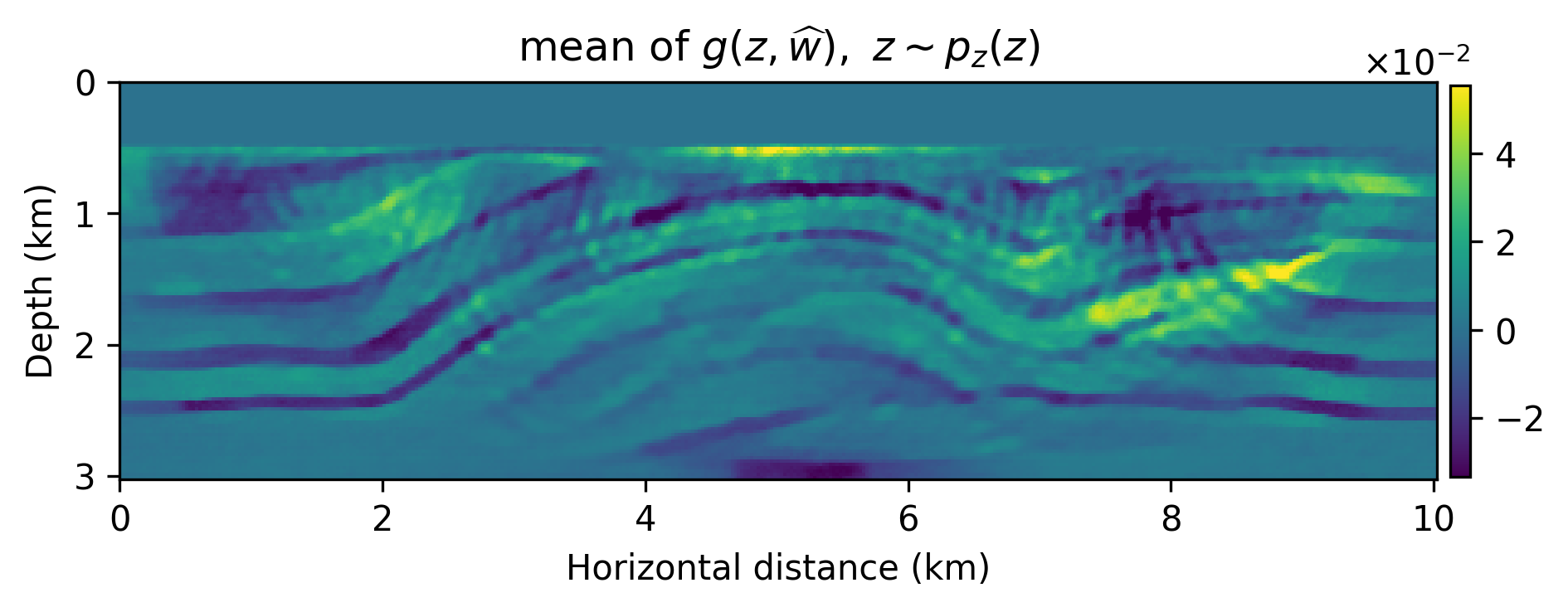}}
\\
\subfloat[\label{figure-1c}]{\includegraphics[width=0.500\hsize]{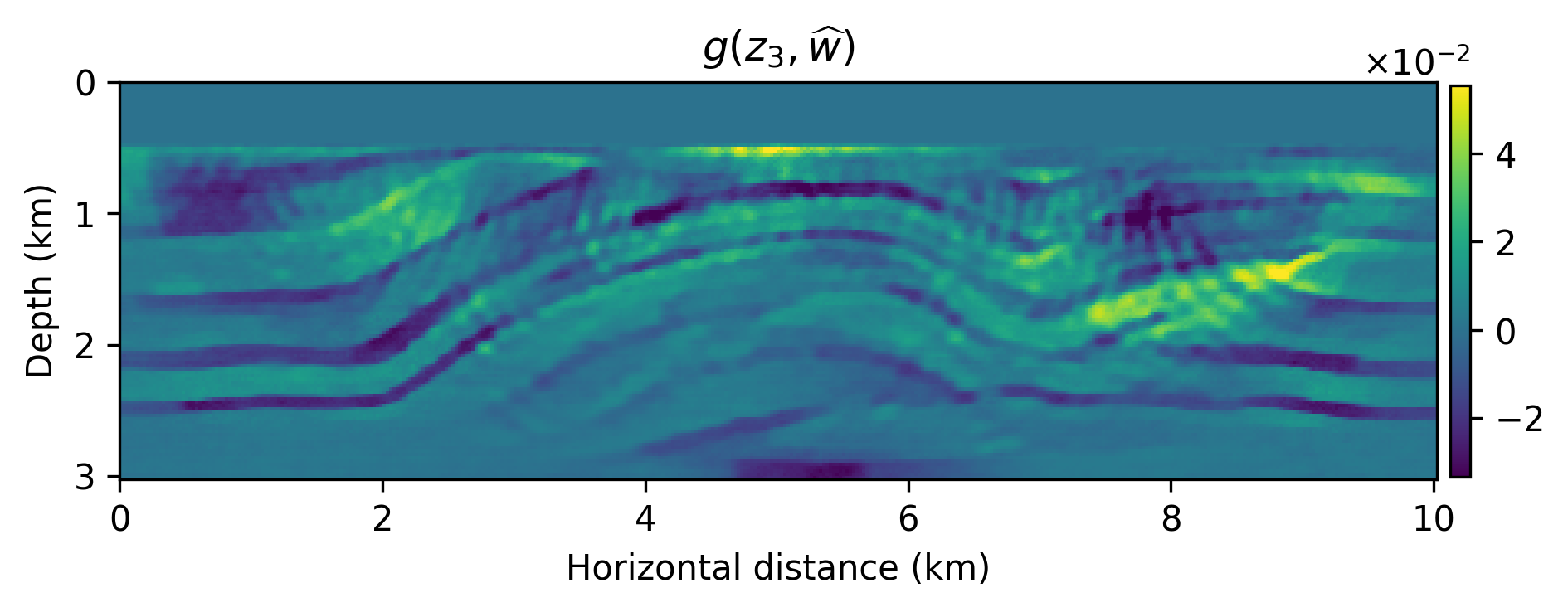}}
\subfloat[\label{figure-1d}]{\includegraphics[width=0.500\hsize]{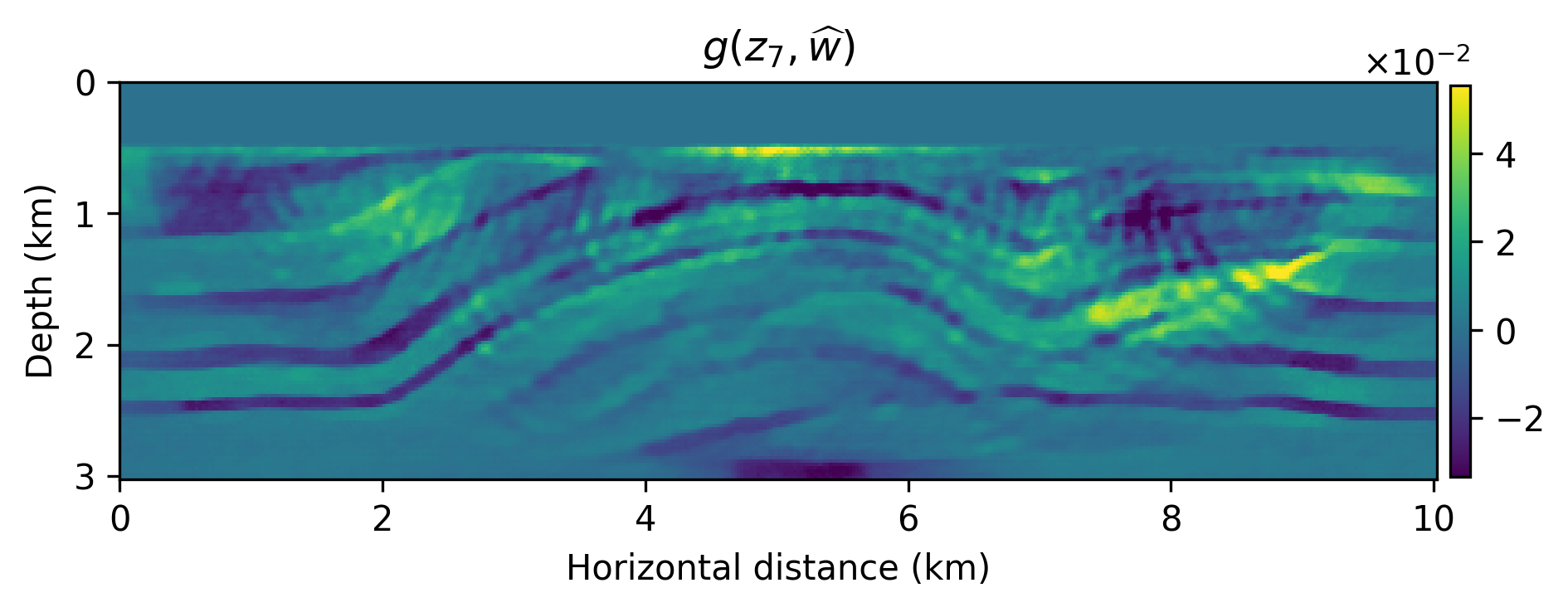}}
\\
\subfloat[\label{figure-1e}]{\includegraphics[width=0.500\hsize]{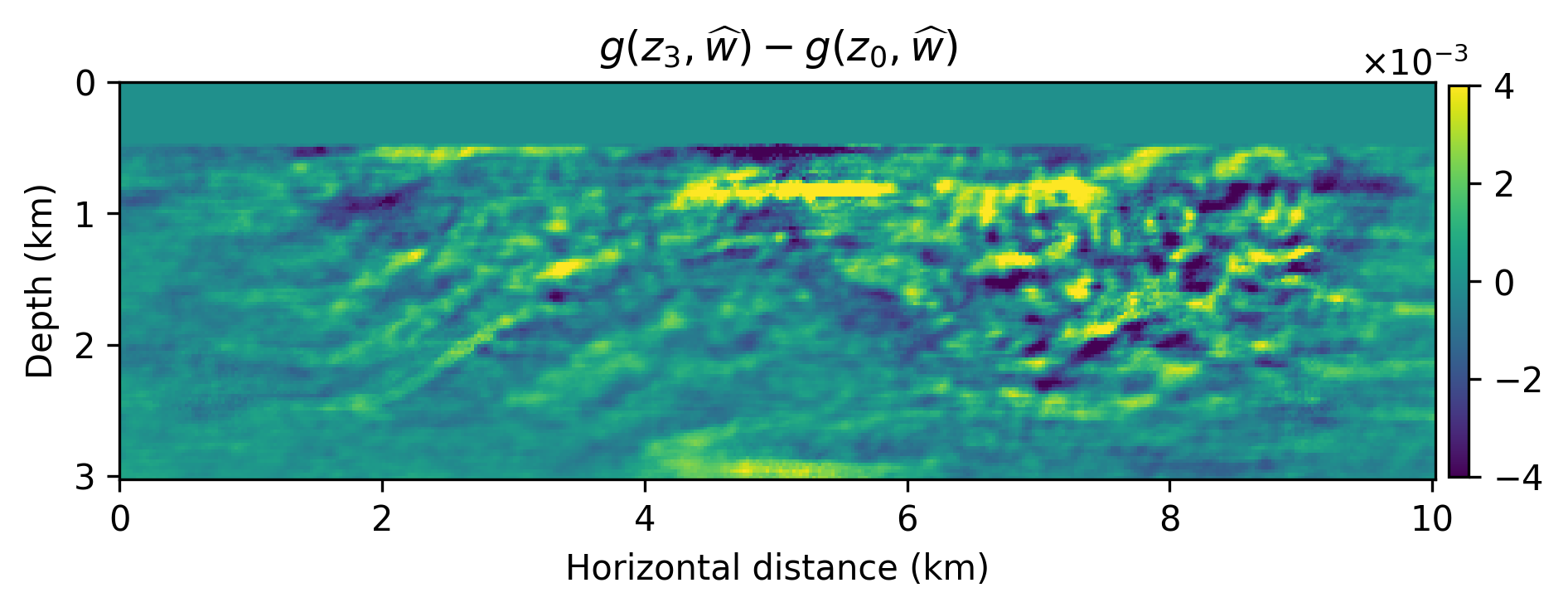}}
\subfloat[\label{figure-1f}]{\includegraphics[width=0.500\hsize]{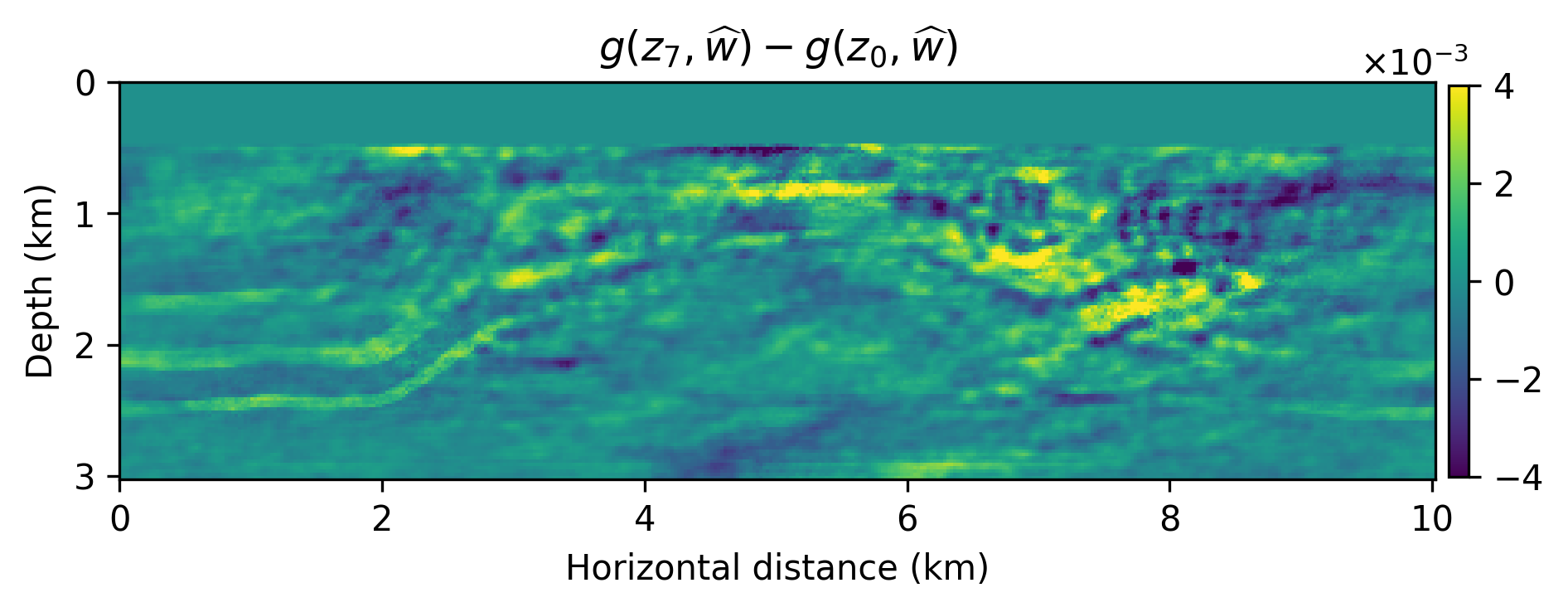}}
\caption{Imaging according to the proposed method. a) a Bregman primal
variable $x_{i}^\ast$ obtained after $350$ Bregman iterations. b) the
mean of $g(z,\widehat{w})$ obtained by generating $3200$ random
realizations of $z\sim p_z(z)$ and averaging the corresponding
$g(z,\widehat{w})$'s. c,d) two examples of generated images from
$g(z,\widehat{w})$ for different $z$'s. e,f) the differences between
images in the middle row with another realization of the
network.}\label{fig:uq}
\end{figure*}

We numerically conduct a survey where the source experiments contain
severe incoherent noise and coherent linearization errors:
$e=(F_k(m+\delta m)-F_k(m)-\nabla F_k(m)\delta m)$, where
$A_k=\nabla F_k$ is the Jacobian and $F_k(m)$ is the nonlinear forward
operator with $m$ the known smooth background model and $\delta m$ the
unknown perturbation (image). The signal-to-noise ratio of the observed
data is $-11.37$ dB. The results of this experiment are included in
Figure~\ref{fig:uq} from which we make the following observations.
First, as expected the models generated from $g(z,\widehat{w})$ are
smoother than the primal Bregman variable. Second, there are clearly
variations amongst the different $g(z,\widehat{w})$'s and these
variations average out in the mean, which has fewer imaging artifacts.

\begin{figure*}
\centering
\subfloat[\label{figure-2a}]{\includegraphics[width=0.650\hsize]{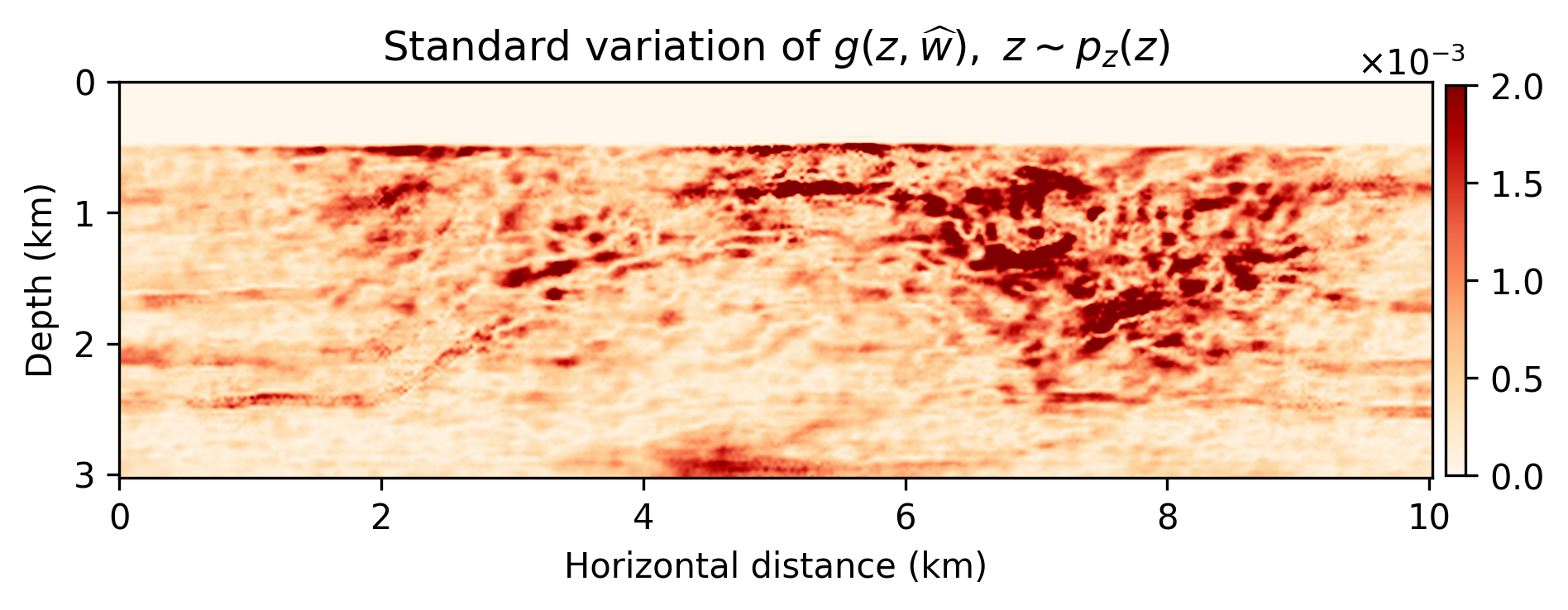}}
\\
\subfloat[\label{figure-2b}]{\includegraphics[width=0.350\hsize]{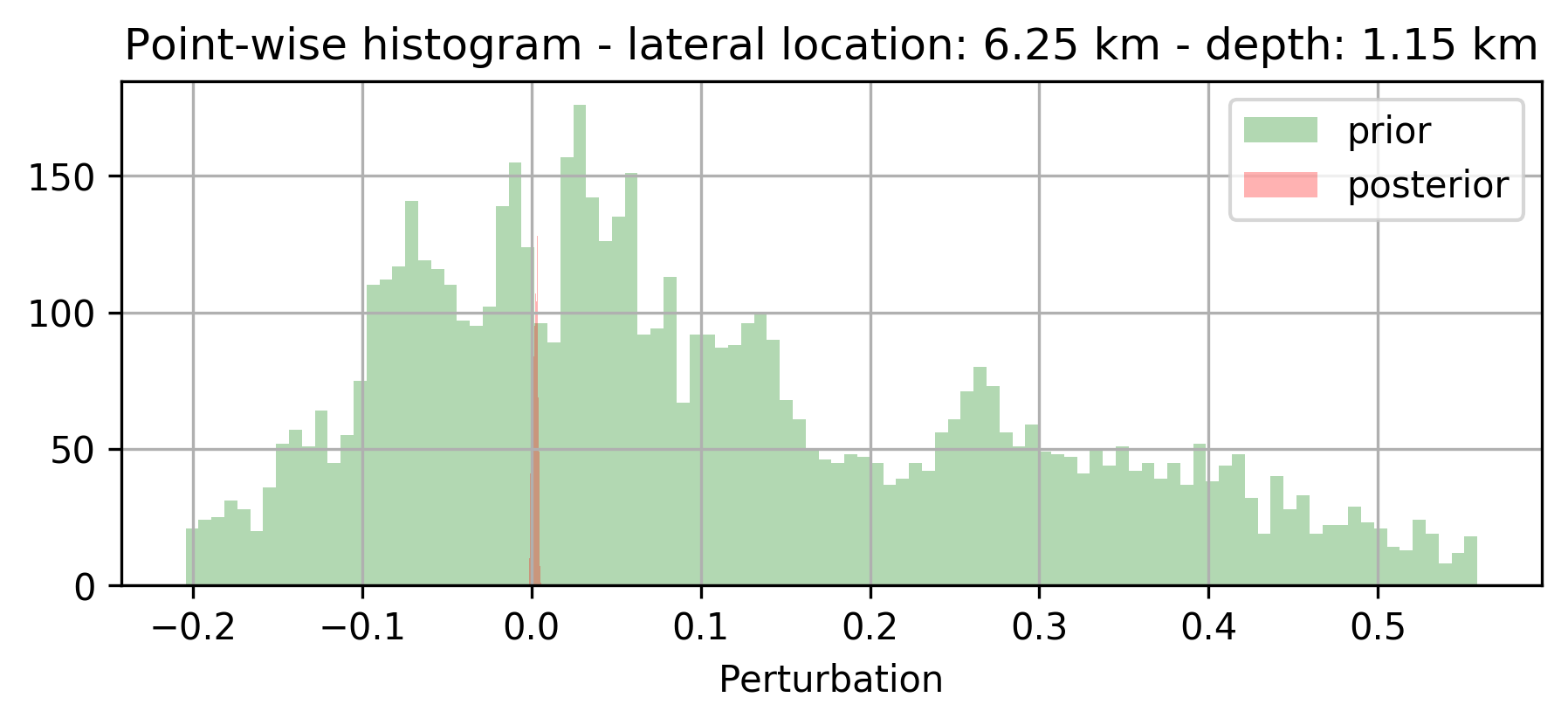}}
\subfloat[\label{figure-2c}]{\includegraphics[width=0.350\hsize]{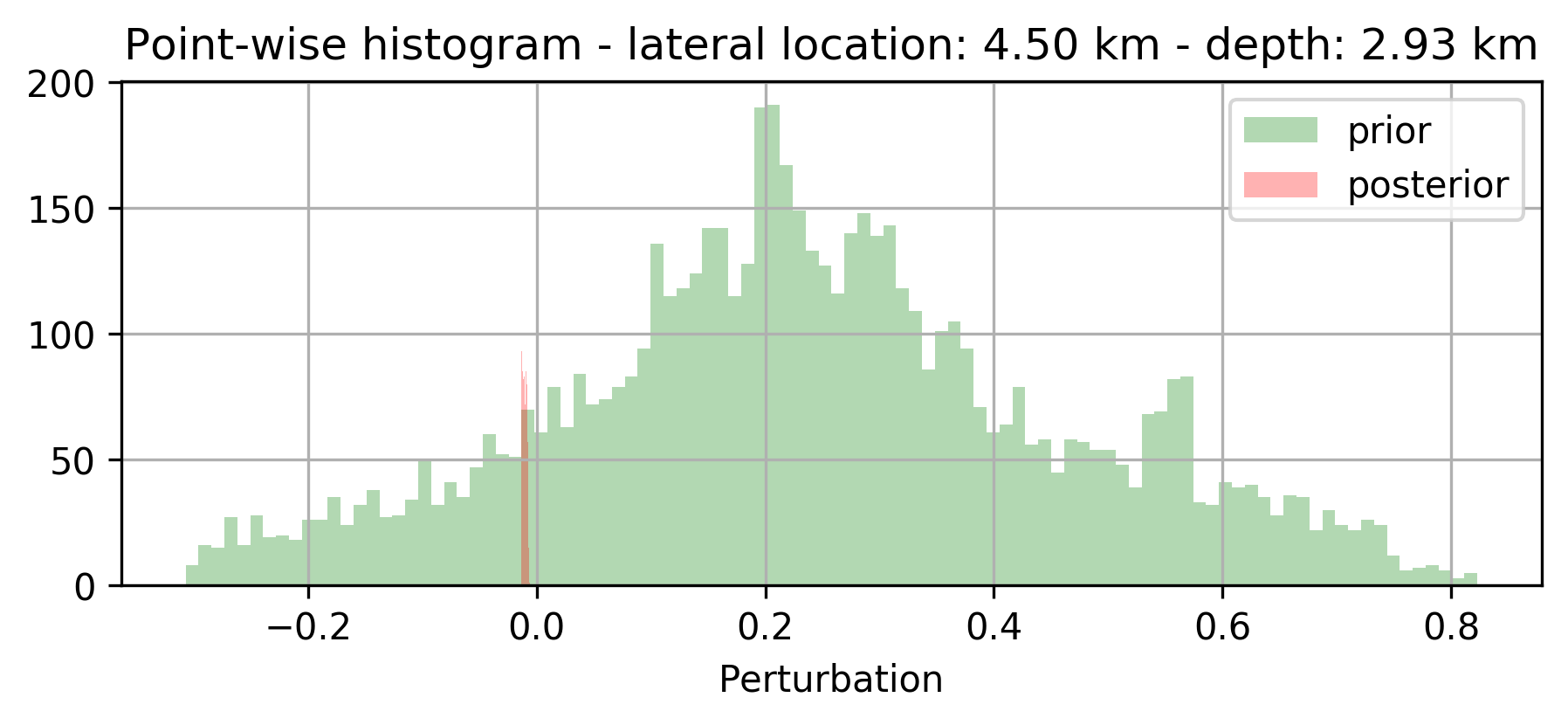}}
\caption{Statistics of imaging according to the proposed method. a) the
pointwise standard deviation among samples generated by evaluating
$g(z,\widehat{w})$ over $3200$ random realizations of
$g(z, w), \ z\sim p_z(z)$. b,c) sample ``prior'' (before training) and
``posterior'' distribution functions for two points in the
model.}\label{fig:stats}
\end{figure*}

Because we were able to train the $g(z,w)$ as a ``byproduct'' of the
inversion, we are able to compute statistical information from the
trained generative model that may give us information about the
``uncertainty''. In Figure~\ref{fig:stats}, we included a plot of the
pointwise standard deviation , computed with $3200$ random realizations
of $g(z, w), \ z\sim p_z(z)$, and two examples of sample ``prior''
(before training) and ``posterior'' distribution. As expected, the
pointwise standard deviations shows a reasonable sharpening of the
probabilities before and after training through inversion. We also argue
that the areas of high pointwise standard deviation coincide with
regions that are difficult to image because of the linearization error
and noise.

\section{Discussion and Conclusions}\label{discussion-and-conclusions}

In this work, we tested an inverse problem framework which includes hard
constraints and deep priors. Hard constraints are necessary in many
problems, such as seismic imaging, where the unknowns must belong to a
feasible set in order to ensure the numerical stability of the forward
problem. Deep priors, enforced through adherence to the range of a
neural network, provide an additional, implicit type of regularization,
as demonstrated by recent work
\citep[\citet{dittmer2018regularization}]{Veen2018}, and corroborated by
our numerical results. The resulting algorithm can be mathematically
interpreted in light of expectation maximization methods. Furthermore,
connections to elastic averaging SGD \citep{zhang2015deep} highlight
potential computational benefits of a parallel (synchronous or
asynchronous) implementation.

On a speculative note, we argue that the presented method, which
combines stochastic optimization on the dual variable with on-the-fly
estimation of the generative model's weights using Langevin dynamics,
reaps information on the ``posterior'' distribution leveraging
multiplicity in the data and the fact that the data is acquired over one
and the same Earth model. Our preliminary results seem consistent with a
behavior to be expected from a ``posterior'' distribution.

\bibliography{learnedsiminv}

\end{document}